\documentclass [preprint]{jpsj2}
\usepackage[usenames]{color}
\usepackage{graphicx}% Include figure files
\usepackage{dcolumn}% Align table columns on decimal point

\def\runtitle{Charge Ordering in $\alpha$-(BEDT-TTF)$_2$I$_3$}

\def\runauthor{T.{\sc Kakiuchi} {\it et al}.}

\title{Charge Ordering in $\alpha$-(BEDT-TTF)$_2$I$_3$ by synchrotron x-ray diffraction}

\author{Toru Kakiuchi$^{1}$, Yusuke Wakabayashi$^{1,2}$, Hiroshi Sawa$^{1,2}$\footnote{e-mail address: hiroshi.sawa@kek.jp},  Toshihiro Takahashi$^{3}$, and Toshikazu Nakamura$^{4}$}

\inst{$^{1}$Department of Materials Structure Science, The Graduate University for Advanced Studies, Tsukuba, Ibaraki 305-0801, Japan
\\$^{2}$Institute of Materials Structure Science, High Energy Accelerator Research Organization, Tsukuba, Ibaraki 305-0801, Japan
\\$^{3}$Department of Physics, Gakushuin University, Mejiro, Tokyo 171-8588, Japan
\\$^{4}$Institute for Molecular Science, Myodaiji, Okazaki 444-8585, Japan}

\recdate{\today}

\abst{ The spatial charge arrangement of a typical quasi-two-dimensional organic conductor $\alpha$-(BEDT-TTF)$_2$I$_3$ is revealed by single crystal structure analysis using synchrotron radiation. The results show that the horizontal stripe type structure, which was suggested by mean field theory, is established. We also find the charge disproportion above the metal-insulator transition temperature and a significant change in transfer integrals caused by the phase transition. Our result elucidates the insulating phase of this material as a 2$k_F$ charge density localization. 
}

\kword{$\alpha$-(BEDT-TTF)$_2$I$_3$, Charge Ordering, Quasi two dimensional organic conductor, Synchrotron Radiation, Crystal structure analysis}

\begin{document}
\sloppy
\maketitle

%#############################################################
As an origin of a metal-insulator (M-I) transition, the charge ordering caused by the long range Coulomb interaction between electrons is a prominent phenomenon as well as the Peierls transition or the Mott Hubbard transition. Organic materials having low-dimensionality provide us a field of studying such interesting phenomena\cite{Fukuyama06JPSJ}.
Among them, an organic conductor $\alpha$-(BEDT-TTF)$_2$I$_3$, which undergoes a M-I transition at $T_{MI}=135$~K\cite{bender_synth}, attracts considerable attention because of its possibility of zero-gap semiconductor under hydrostatic pressures\cite{tajima06JPSJ}. 
The insulating phase is a nonmagnetic state with a spin gap\cite{Rothaemel86PRB}, and interpreted to be caused by a charge disproportionation theoretically\cite{kino_fukuyama-1,seo} and experimentally\cite{nmr-1,raman}. According to these studies, horizontal-charge stripes with valences of +1$e$ and 0 are expected on BEDT-TTF molecules while no structural evidence has been reported.
 In this letter, we report a precise structure of $\alpha$-(BEDT-TTF)$_2$I$_3$ as a function of temperature based on synchrotron radiation diffraction measurements in order to provide evidence of the charge disproportionation.

$\alpha$-(BEDT-TTF)$_2$I$_3$, which has a two-dimensional (2D) conduction band, is a typical organic conductor made of I$_3^-$ ions and BEDT-TTF$^{0.5+}$ on average. The crystal structure of this material is a sandwich structure of I$_3$ insulating layers and 2D-BEDT-TTF conduction layers having a quarter-filled hole band. The space group at room temperature is $P\bar{1}$. Together with the metal-insulator transition, it shows paramagnetic-nonmagnetic transition\cite{Rothaemel86PRB}. According to several reports\cite{hinetsu,kino_fukuyama-2,seo}, this transition is not a Peierls transition but a charge ordering caused by a strong Coulomb repulsion. 
Theoretically, the charge ordering of this material was first discussed by Kino and Fukuyama with the Hubbard model\cite{kino_fukuyama-2}, and Seo performed extended Hubbard model calculation\cite{seo} that predicted several arrangements of charge ordering is stabilized as a function of $U/V$, where $U$ and $V$ are on-site and inter-site Coulomb energy. 
 Experimentally, a so-called ``horizontal stripe" type of charge ordering is suggested by NMR\cite{nmr-1} and Raman scattering\cite{raman} measurements. However, no structural evidence for this charge ordering has been provided despite of several attempts\cite{xray-endres,xray-emge,xray-nogami}. 

 Single crystals of $\alpha$-(BEDT-TTF)$_2$I$_3$ were grown using an electrochemical method. The shape of the specimens used for our measurements was a quadratic prism with a typical edge-length of 0.1mm. Oscillation photographs with 18~keV x-ray for crystal structure analysis were taken with an imaging plate (IP) Weissenberg camera installed to beam lines BL-1A and -1B at the Photon Factory, KEK, Japan. The sample temperature was controlled by a N$_2$ flow cooler (down to 90~K) and a closed-cycle helium refrigerator (down to 20~K).  For image data processing of digital IP data and structure refinement, RAPID software and CrystalStrucuture software (RIGAKU Co, Ltd.) were used. Additionally, we used a four-circle diffractometer having scintillation counter installed to BL-4C for symmetry determination based on the Friedel law.
 The x-ray energy for this measurement was 12.4KeV in order to utilize a large value of anomalous scattering factor, and the N$_2$ flow cooler was used for controlling the sample temperature.

\begin{figure}
\includegraphics[scale = 0.4]{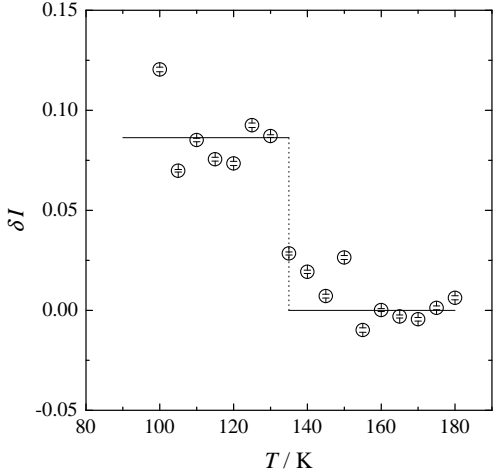}
\caption{Intensity difference $\delta I$ of ($\bar{2},\bar{8},6$) and ($2,8,\bar{6}$). 
The abrupt change at $T_{MI}$ shows the breaking of the inversion symmetry. Solid lines are guides to the eyes. 
 }
\label{friedel}
\end{figure}

Utilizing the Friedel law, the intensity of the Bragg reflection of the Miller index $(h,k,l)$ ($I(h,k,l)$) equals $I(\bar{h},\bar{k},\bar{l})$ for centrosymmetric crystals, we firstly examined the symmetry reduction from $P\bar{1}$ to $P1$ accompanied by the M-I transition, which was suggested by Raman scattering measurements\cite{raman}. 
Figure \ref{friedel} shows the intensity difference $\delta I \equiv \frac{I(h,k,l)-I(\bar{h},\bar{k},\bar{l})}{I(h,k,l)+I(\bar{h},\bar{k},\bar{l})}$ for ($\bar{2},\bar{8},6$) and ($2,8,\bar{6}$) as a function of temperature. The absorption correction was made using the data for higher temperatures (165K - 180K) at which the crystal is centrosymmetric. 
The jump in the value of $\delta I$ at $T_{\textrm{MI}}$ seen in Fig. \ref{friedel} indicates the breaking of inversion symmetry at the transition temperature. Therefore, the space group of insulating phase was found to be $P1$ uniquely. 

In order to make clear the charge ordering structure, we performed crystal structure analyses at room temperature (RT) and 20K. Lattice parameters at RT  were $a=9.1868(13)$\AA, $b=10.7926(18)$\AA, $c=17.400(3)$\AA, $\alpha=96.957(4)^{\circ}$, $\beta=97.911(6)^{\circ}$, $\gamma=90.795(6)^{\circ}$ and $V=1695.4(5)$\AA$^3$, 
and at 20K were $a=9.0162(4)$\AA, $b=10.6695(6)$\AA, $c=17.324(1)$\AA, $\alpha=96.536(3)^{\circ}$, $\beta=97.758(4)^{\circ}$, $\gamma=90.198(4)^{\circ}$ and $V=1639.5(1)$\AA$^3$. The resolution of the data and the number of unique reflections at RT (20K) are 0.48\AA (0.40\AA) and 16708 (39714). The structure refinement using the full matrix least square method was converged well, showing a good reliability factor $R_{\textrm{all}}=5.2\%$(RT) and 5.6\%(20K). In the case of an acentric structure, the crystal has a twin structure of the right-handed coordinated system and the left-handed one, and the Flack parameter, which is an indicator of volume fraction between them, was 0.3 at 20~K, meaning that our data were taken from a crystal having 70\% of the right-handed domain and 30\% of the left-handed domain. The detailed structural parameters are 
shown in Table~\ref{tbl:list}.\cite{Kakiuchi_DB}
The detail of the twin structure is discussed later.
\begin{table*}[htb]
\caption{\label{strparam}Refined fractional coordinates and isotropic displacement parameters at 20K. }
\label{tbl:list}
{\small
\hspace{-1cm}
\begin{tabular}{cccccccccc}
\hline
 atom  &  $x$  &  $y$  &  $z$  &  $B_{\textrm{eq}}$ & atom  &  $x$  &  $y$  &  $z$  &  $B_{\textrm{eq}}$ \\
\hline
I1& 0.804358(12)&-0.07710(2)& 1.015100(8)& 0.315(3)&      C2& 0.2472(2)& 0.2849(4)& 0.82515(15)& 0.44(6)\\
I2& 0.183265(12)& 0.07630(2)& 0.989830(8)& 0.297(3)&      C3& 0.3175(2)& 0.0942(4)& 0.68850(14)& 0.28(6)\\
I3& 0.689448(13)& 0.41917(2)& 1.007854(8)& 0.296(3)&      C4& 0.2288(2)& 0.1893(4)& 0.66858(14)& 0.26(6)\\
I4& 0.493555(19)&-0.00072(3)& 1.001866(11)& 0.196(2)&     C5& 0.2379(2)& 0.0469(3)& 0.53856(13)& 0.26(6)\\
I5& 1.310619(13)& 0.57238(2)& 0.994637(8)& 0.323(3)&      C6& 0.2135(2)&-0.0099(4)& 0.46352(14)& 0.31(6)\\
I6& 1.00002(2)& 0.49570(3)& 1.000934(12)& 0.207(2)&       C7& 0.2175(2)&-0.1594(4)& 0.33563(14)& 0.28(6)\\
S7& 0.59349(7)&-0.07374(11)& 0.22174(4)& 0.278(17)&       C8& 0.1287(2)&-0.0664(4)& 0.31442(14)& 0.29(6)\\
S8& 0.81283(7)& 0.19331(10)& 0.27851(4)& 0.270(16)&       C9& 0.1683(2)&-0.2776(4)& 0.18406(15)& 0.38(6)\\
S9& 0.88578(7)&-0.06581(10)& 0.61617(4)& 0.261(16)&       C10& 0.1649(2)&-0.1438(4)& 0.16341(15)& 0.49(7)\\
S10& 0.29722(7)&-0.15077(11)& 0.43456(4)& 0.272(17)&      C11& 0.8232(2)& 0.1401(4)& 0.83927(15)& 0.36(6)\\
S11& 1.18076(7)& 0.80633(10)& 0.72772(4)& 0.269(16)&      C12& 0.8213(2)& 0.2730(4)& 0.81703(15)& 0.45(7)\\
S12& 0.70010(7)& 0.14045(11)& 0.56717(4)& 0.277(17)&      C13& 0.8592(2)& 0.0553(4)& 0.68924(14)& 0.27(6)\\
S13& 0.34721(7)& 0.18209(11)& 0.27781(4)& 0.311(16)&      C14& 0.7732(2)& 0.1517(4)& 0.66588(14)& 0.27(6)\\
S14& 0.82946(7)&-0.20140(10)& 0.43355(4)& 0.271(16)&      C15& 0.7788(2)&-0.0015(4)& 0.54062(13)& 0.29(6)\\
S15& 0.10013(7)& 0.05485(10)& 0.38875(4)& 0.266(16)&      C16& 0.7548(2)&-0.0592(4)& 0.46424(14)& 0.33(6)\\
S16& 0.57971(7)& 0.67695(11)& 0.56781(4)& 0.278(17)&      C17& 0.7595(2)&-0.2009(4)& 0.33445(14)& 0.31(6)\\
S17& 0.27178(7)&-0.29284(10)& 0.27918(4)& 0.287(17)&      C18& 0.6742(2)&-0.1026(4)& 0.31499(14)& 0.27(6)\\
S18& 0.40699(7)& 0.32078(11)& 0.43462(4)& 0.309(17)&      C19& 0.5883(2)&-0.2316(4)& 0.17023(15)& 0.33(7)\\
S19& 0.40028(7)& 0.06939(11)& 0.78259(4)& 0.278(17)&      C20& 0.7387(2)&-0.2924(4)& 0.17711(14)& 0.32(6)\\
S20& 0.64326(7)& 0.00906(11)& 0.39221(4)& 0.266(16)&      C21& 1.0051(2)& 0.7202(4)& 0.83316(15)& 0.35(6)\\
S21& 0.80468(7)&-0.33318(11)& 0.27450(4)& 0.289(16)&      C22& 1.1617(2)& 0.7688(4)& 0.82548(14)& 0.35(6)\\
S22& 0.35638(7)&-0.01732(10)& 0.61164(4)& 0.253(16)&      C23& 0.9989(2)& 0.5856(4)& 0.68816(14)& 0.29(6)\\
S23& 1.11533(7)& 0.67020(10)& 0.56957(4)& 0.255(15)&      C24& 1.0891(2)& 0.6756(4)& 0.66870(14)& 0.26(6)\\
S24& 0.15846(7)& 0.18908(10)& 0.56894(4)& 0.274(16)&      C25& 1.0029(2)& 0.5324(4)& 0.53886(14)& 0.30(6)\\
S25& 0.96013(7)& 0.56331(11)& 0.78213(4)& 0.297(16)&      C26& 0.9864(2)& 0.4739(4)& 0.46423(13)& 0.28(6)\\
S26& 1.03670(7)& 0.43202(11)& 0.22159(4)& 0.307(17)&      C27& 0.9033(2)& 0.3258(4)& 0.33614(14)& 0.29(6)\\
S27& 0.43894(7)& 0.45180(11)& 0.61318(4)& 0.310(16)&      C28& 0.9947(2)& 0.4134(4)& 0.31565(14)& 0.26(6)\\
S28& 0.51093(7)& 0.45378(11)& 0.21839(4)& 0.340(16)&      C29& 0.8278(2)& 0.2318(4)& 0.18086(15)& 0.39(6)\\
S29& 0.63810(7)& 0.81348(11)& 0.72524(4)& 0.309(16)&      C30& 0.9839(2)& 0.2766(4)& 0.17096(15)& 0.40(6)\\
S30& 0.04441(7)&-0.04617(11)& 0.21948(4)& 0.280(16)&      C31& 0.5973(2)& 0.6611(4)& 0.84206(15)& 0.34(6)\\
S31& 0.17758(7)& 0.32066(10)& 0.72748(4)& 0.283(16)&      C32& 0.5766(2)& 0.7919(4)& 0.81816(15)& 0.48(7)\\
S32& 0.54364(7)& 0.54632(11)& 0.38784(4)& 0.274(16)&      C33& 0.4982(2)& 0.5725(4)& 0.68854(14)& 0.30(6)\\
S33& 1.07593(7)& 0.53098(10)& 0.39066(4)& 0.249(16)&      C34& 0.5629(2)& 0.6766(4)& 0.66683(14)& 0.30(6)\\
S34& 0.94100(7)& 0.03797(11)& 0.78465(4)& 0.280(16)&      C35& 0.4976(2)& 0.5280(4)& 0.53892(14)& 0.32(6)\\
S35& 0.87132(7)& 0.33820(10)& 0.43441(4)& 0.291(16)&      C36& 0.4833(2)& 0.4710(4)& 0.46233(13)& 0.38(6)\\
S36& 0.91127(7)& 0.47539(10)& 0.61144(4)& 0.262(16)&      C37& 0.4199(2)& 0.3196(4)& 0.33504(14)& 0.33(6)\\
S37& 0.71854(7)& 0.28560(10)& 0.72114(4)& 0.289(16)&      C38& 0.4827(2)& 0.4242(4)& 0.31294(14)& 0.34(6)\\
S38& 0.46812(7)& 0.54336(11)& 0.78248(4)& 0.329(16)&      C39& 0.4161(2)& 0.2006(4)& 0.18632(15)& 0.34(6)\\
C1& 0.4013(2)& 0.2284(4)& 0.83231(15)& 0.34(7)&           C40& 0.3921(2)& 0.3295(4)& 0.15897(15)& 0.47(7)\\
\hline

\end{tabular} 
}
\end{table*}

\begin{figure}
\includegraphics[scale = 0.5]{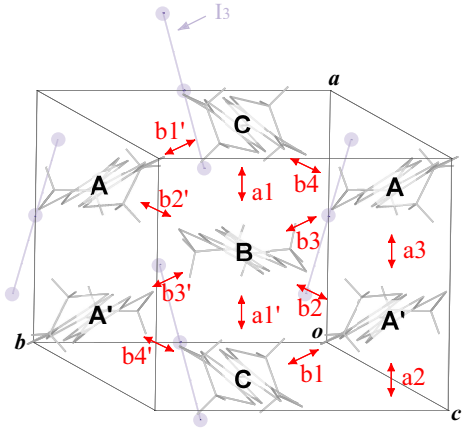}
\caption{Crystal structure of $\alpha$-(BEDT-TTF)$_2$I$_3$ seen from the molecular longitudinal direction.
Each molecule (A, A$'$, B and C) is independent crystallographically below $T_{\textrm{MI}}$ 
although A and A$'$ are equivalent above $T_{MI}$ because of the inversion symmetry. 
The double-side arrows show the labels for the overlap integrals. 
 }
%2
\label{structure}
\end{figure}

\begin{figure}
\includegraphics[scale = 0.4]{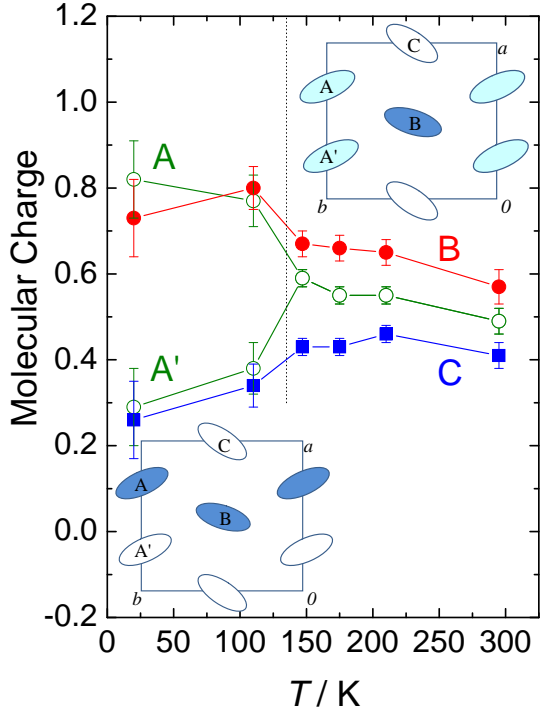}
\caption{Temperature dependence of the molecular charges. 
The charge disproportion enlarges significantly at $T_{\textrm{MI}}$. 
 }
%3
\label{charge}
\end{figure}

Figure \ref{structure} shows the crystal structure of $\alpha$-(BEDT-TTF)$_2$I$_3$ seen from molecular longitudinal direction. At room temperature, there are three crystallographically independent molecules, called A, B and C, in a unit cell. 
Molecule A$'$ is equivalent to A owing to the inversion symmetry. The valence of a BEDT-TTF molecule is estimated by an empirical method based on the bond lengths in the molecule\cite{guionneau}. The result of the structure refinement shows the valence conditions of A=0.49(3), B=0.57(4) and C=0.41(3) at RT (see upper inset to Fig.\ref{charge}).
This charge disproportion in the metallic state cannot be caused by the local electric potential distribution originated from the I$_3^-$ ions, because the valence of the BEDT-TTF molecule that is the closest of I$_3^-$ ion is 0.4+, rather more negatively charged than average. Therefore, the charge disproportion has something to do with the transfer integrals among the BEDT-TTF molecules. It is noteworthy that the weak charge disproportion was also observed at RT as reported previously \cite{raman,ishibashi,moroto}.

In the low-temperature insulating phase, we have succeeded in a direct determination of the charge ordering structure using the same method as above. The space group is $P1$ and there are four independent molecules in the unit cell, or molecule A$'$ is independent on A. We observed molecular deformations, which reflect charge disproportion, at 20K. Estimated each charge value is A=0.82(9), A$'$=0.29(9), B=0.73(9) and C=0.26(9).  The charge structure is a horizontal stripe type, in which molecules A and B are hole rich and A$'$ and C are hole poor sites as shown in the lower inset to Fig.\ref{charge}. The obtained arrangement corresponds to the theoretical predicted one by Seo\cite{seo}.

\begin{figure}
\includegraphics[scale = 0.4]{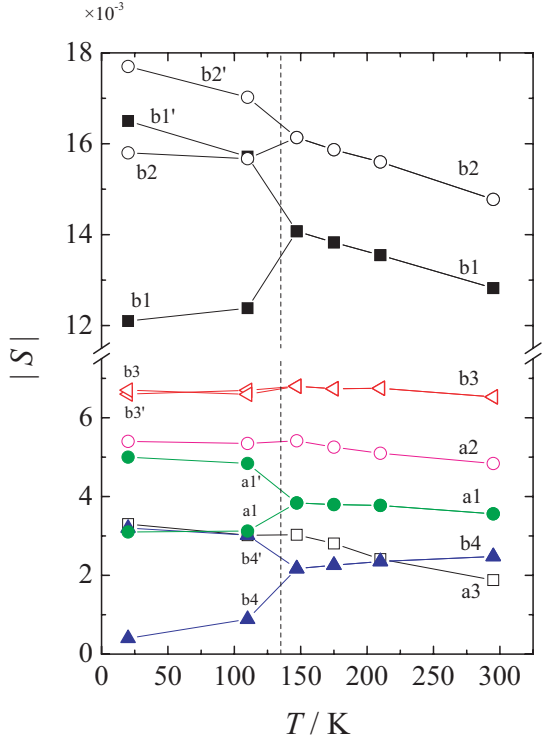}
\caption{Temperature dependence of the overlap integrals. 
Because of symmetry reduction from $P\bar{1}$ to $P1$, there are twelve integrals 
below $T_{\textrm{MI}}$ although seven integrals in metallic phase. 
All labels correspond to those in Fig. \ref{structure}. 
 }
%4
\label{overlap}
\end{figure}

To investigate the electronic and structural change associated with the charge ordering transition, we performed the structure analyses at intermediate temperatures, 210K, 175K, 147K and 110K. The reliability factors of structure refinements are small enough (ca. 5\%) for making a precise discussion of electronic state. Figure\ref{charge} shows the temperature dependence of the molecular charges. The relation of the molecular charge (B$>$A$>$C) observed at RT conserves in the metallic state, while the charge distribution changes gradually. Based on the tight binding approximation and a molecular orbital calculation with the extended Huckel method\cite{mo_calc}, we calculated the overlap integrals $|S|$'s between the neighboring molecules, which are proportional to the transfer integrals approximately. Figure \ref{overlap} shows the temperature dependence of the overlap integrals.
 Labels in the figure correspond to those of Fig. \ref{structure}, in which the $|S|$'s having labels with and without an apostrophe are equivalent in metallic phase because of the inversion symmetry. The number of the overlap integrals increases from seven in the metallic phase to twelve in the insulating phase due to the reduction of symmetry.

\begin{figure}
\includegraphics[width=7cm]{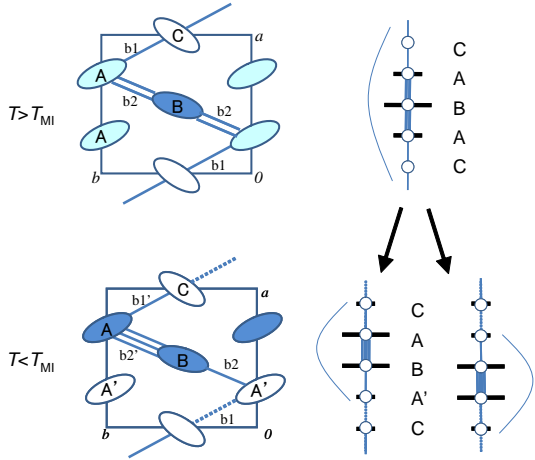}
\caption{(left) Schematic view of a zigzag chain made of large overlap integrals $|S|$b1 and $|S|$b2. 
Thickness of the line shows the magnitude of the overlap integrals. 
(right) Expanded picture from zigzag chain to a one dimensional chain. 
In the low temperature phase, there are two possibilities of the valence arrangements to take. 
 }
%5
\label{schematic}
\end{figure}

It is noteworthy that the overlap integrals are able to be classified into two groups; b1 and b2 form a main transfer path regarded as a one dimensional ``zigzag'' chain in the unit cell and the other paths form inter-chain interaction as shown in Fig.\ref{schematic}(upper).
 The values of $|S|$'s for b1 and b2 in the chain increase with decreasing temperature down to $T_{MI}$. 

Below $T_{MI}$, overlap integral $|S|$b1 which corresponds to the hopping from molecule A$'$ to molecule C become fairly small and the values of $|S|$b2 and $|S|$b1$'$ approach each other. The modulated transfer integrals in a zigzag chain are schematically shown in Fig. \ref{schematic}(lower). The structure is very similar to the 2$k_F$ charge ordering, where $k_F$ is the Fermi wave number. This 2$k_F$ structure causes both charge gap and spin gap, and makes the system a nonmagnetic spin singlet state. From these results, we can recognize that the electronic (band) structure exhibit an abrupt change accompanied by M-I transition. 

The main structural change creating this modulation of transfer integrals is not translational shifts of molecule but  changes in the dihedral angles. The maximum degree of change in the angles between RT and 20K was about 3$^\circ$, and the change in the angle make the spatial overlap of the wavefunctions. This structural charge ordering pattern would be described by the ordinary low-dimensional electronic state and is in striking contrast to Wigner crystallization of the charge ordering in quarter filled conductor, (DI-DCNQI)$_2$Ag system\cite{prl_hiraki,prl_kakiuchi}.

Finally, we mention the twin structure. When the crystal undergoes the phase transition from $P\bar1$ to $P1$, there are two kinds of structural deformation. The situation for this particular case is depicted in Fig.\ref{schematic}. One side of the twin is the (A and B)-rich type and the other is (A$'$ and B)-rich type. The left-side panel for the low-temperature phase in this figure shows the former. The overlap integrals for b1$'$ and b2$'$ are larger than those for b1 and b2, respectively, in the former, and the other way around in the latter. The differences between them are only the phase of charge density, and the electronic states of them are completely equivalent. 

We succeeded in the determination of the charge arrangement in the insulating phase of $\alpha$-(BEDT-TTF)$_2$I$_3$ by means of synchrotron radiation x-ray crystal structure analysis. Based on the molecular deformation, we found horizontal type of charge ordering directly. In the metallic phase, charge disproportion in the conduction plane was also observed. The result of a quantum chemical calculation based on the obtained structure parameters shows a drastic change in transfer integrals accompanied by M-I transition. The insulator state is described with 2$k_F$ charge density wave within a zigzag path connected by the large transfer integrals.

 Authors are grateful to Dr. H. Seo for his fruitful discussion. This work was partially supported by a Grant-in-Aid for Creative Scientific Research and Scientific Research from the Ministry of Education, Culture, Sports, Science, and Technology of Japan.

%REFERENCE

\end{document}